\documentclass[twocolappendix,appendixfloats]{emulateapj}

\usepackage[colorlinks = true, linkcolor = blue, urlcolor  = blue, citecolor = blue, anchorcolor = blue]{hyperref}
\usepackage{hyperref}
\usepackage{apjfonts}
\usepackage{rotating}
\usepackage{amsmath}
\usepackage{color}
\usepackage{color}
\usepackage{CJK}

\newcommand{\pivec}{\mbox{\boldmath $\pi$}}
\newcommand{\muvec}{\mbox{\boldmath $\mu$}}
\newcommand{\te}{t_{\rm E}}
\newcommand{\thetae}{\theta_{\rm E}}
\newcommand{\pie}{\pi_{\rm E}}
\newcommand{\pien}{\pi_{{\rm E},N}}
\newcommand{\piee}{\pi_{{\rm E},E}}
\newcommand{\dl}{D_{\rm L}}

\definecolor{darkbrown}{RGB}{139,69,19}

\shorttitle{OGLE-2017-BLG-0537}
\shortauthors{JUNG ET AL.}

\begin{document}

\title{OGLE-2017-BLG-0537: Microlensing Event with a Resolvable Lens in $\lesssim 5$ years
from High-resolution Follow-up Observations}

\author{
Y.~K.~Jung\altaffilmark{001,100},
C.~Han\altaffilmark{002,103}, 
A.~Udalski\altaffilmark{003,101}, 
A.~Gould\altaffilmark{001,004,005,100}, \\
and\\
M.~D.~Albrow\altaffilmark{006}, 
S.-J.~Chung\altaffilmark{001,007}, 
K.-H.~Hwang\altaffilmark{001}, 
C.-U.~Lee\altaffilmark{001},
Y.-H.~Ryu\altaffilmark{001},
I.-G.~Shin\altaffilmark{008},
Y.~Shvartzvald\altaffilmark{009,102},
J. C.~Yee\altaffilmark{008},
W.~Zang\altaffilmark{010,011},
W.~Zhu\altaffilmark{012},
S.-M.~Cha\altaffilmark{001,013}, 
D.-J.~Kim\altaffilmark{001}, 
H.-W.~Kim\altaffilmark{001}, 
S.-L.~Kim\altaffilmark{001,007}, 
D.-J.~Lee\altaffilmark{001},
Y.~Lee\altaffilmark{001,013}, 
B.-G.~Park\altaffilmark{001,007}, 
R.~W.~Pogge\altaffilmark{004},
W.-T.~Kim\altaffilmark{014} \\ 
(The KMTNet Collaboration),\\
P.~Mr\'oz\altaffilmark{003}, 
R.~Poleski\altaffilmark{003,004}, 
J.~Skowron\altaffilmark{003}, 
M.~K.~Szyma\'nski\altaffilmark{003}, 
I.~Soszy\'nski\altaffilmark{003},
S.~Koz{\l}owski\altaffilmark{003}, 
P.~Pietrukowicz\altaffilmark{003}, 
K.~Ulaczyk\altaffilmark{003}, 
M.~Pawlak\altaffilmark{003} \\
(The OGLE Collaboration) \\   
}

\email{cheongho@astroph.chungbuk.ac.kr}

\altaffiltext{001}{Korea Astronomy and Space Science Institute, Daejon 34055, Republic of Korea} 
\altaffiltext{002}{Department of Physics, Chungbuk National University, Cheongju 28644, Republic of Korea} 
\altaffiltext{003}{Warsaw University Observatory, Al.~Ujazdowskie 4, 00-478 Warszawa, Poland} 
\altaffiltext{004}{Department of Astronomy, Ohio State University, 140 W. 18th Ave., Columbus, OH 43210, USA} 
\altaffiltext{005}{Max Planck Institute for Astronomy, K\"onigstuhl 17, D-69117 Heidelberg, Germany} 
\altaffiltext{006}{University of Canterbury, Department of Physics and Astronomy, Private Bag 4800, 
                   Christchurch 8020, New Zealand} 
\altaffiltext{007}{Korea University of Science and Technology, 217 Gajeong-ro, Yuseong-gu, Daejeon, 
                   34113, Republic of Korea} 
\altaffiltext{008}{Harvard-Smithsonian Center for Astrophysics, 60 Garden St., Cambridge, MA 02138, USA} 
\altaffiltext{009}{Jet Propulsion Laboratory, California Institute of Technology, 4800 Oak Grove Drive, 
                    Pasadena, CA 91109, USA} 
\altaffiltext{010}{Physics Department and Tsinghua Centre for Astrophysics, Tsinghua University, 
                   Beijing 100084, China}
\altaffiltext{011}{Department of Physics, Zhejiang University, Hangzhou, 310058, China}
\altaffiltext{012}{Canadian Institute for Theoretical Astrophysics, University of Toronto, 60 St George 
                   Street, Toronto, ON M5S 3H8, Canada}
\altaffiltext{013}{School of Space Research, Kyung Hee University, Yongin, Kyeonggi 17104, Korea} 
\altaffiltext{014}{Department of Physics \& Astronomy, Seoul National University, Seoul 08826, Republic of Korea}
\altaffiltext{100}{KMTNet Collaboration.}
\altaffiltext{101}{OGLE Collaboration.}

\altaffiltext{102}{NASA Postdoctoral Program Fellow.}
\altaffiltext{103}{Corresponding author.}

\begin{abstract}
We present the analysis of the binary-lens microlensing event OGLE-2017-BLG-0537.  
The light curve of the event exhibits two strong caustic-crossing spikes among 
which the second caustic crossing was resolved by high-cadence surveys.  It is 
found that the lens components with a mass ratio $\sim 0.5$ are separated in 
projection by $\sim 1.3\thetae$, where $\thetae$ is the angular Einstein radius.  
Analysis of the caustic-crossing part yields $\thetae=1.77\pm 0.16$~mas and a 
lens-source relative proper motion of 
$\mu =12.4\pm 1.1~{\rm mas}~{\rm yr}^{-1}$. The measured $\mu$ is the third highest 
value among the events with measured proper motions and $\sim 3$ times higher than 
the value of typical Galactic bulge events, making the event a strong candidate for 
follow-up observations to directly image the lens by separating it from the source.  
From the angular Einstein radius combined with the microlens parallax, it is estimated 
that  the lens is composed of two main-sequence stars with masses $M_1\sim 0.4~M_\odot$ 
and $M_2\sim 0.2~M_\odot$ located at a distance of $D_{\rm L}\sim 1.2$~kpc.  However, 
the physical lens parameters are not very secure due to the weak microlens-parallax 
signal, and thus we cross check the parameters by conducting a Bayesian 
analysis based on the measured Einstein radius and event timescale combined with the 
blending constraint.  From this, we find that the physical parameters estimated from 
the Bayesian analysis are consistent with those based on the measured microlens parallax.  
Resolving the lens from the source can be done in about 5 years from high-resolution 
follow-up observations and this will provide a rare opportunity to test and refine 
the microlensing model.
\end{abstract}


\section{Introduction}\label{sec:one}

\begin{figure*}
\epsscale{0.88}
\plotone{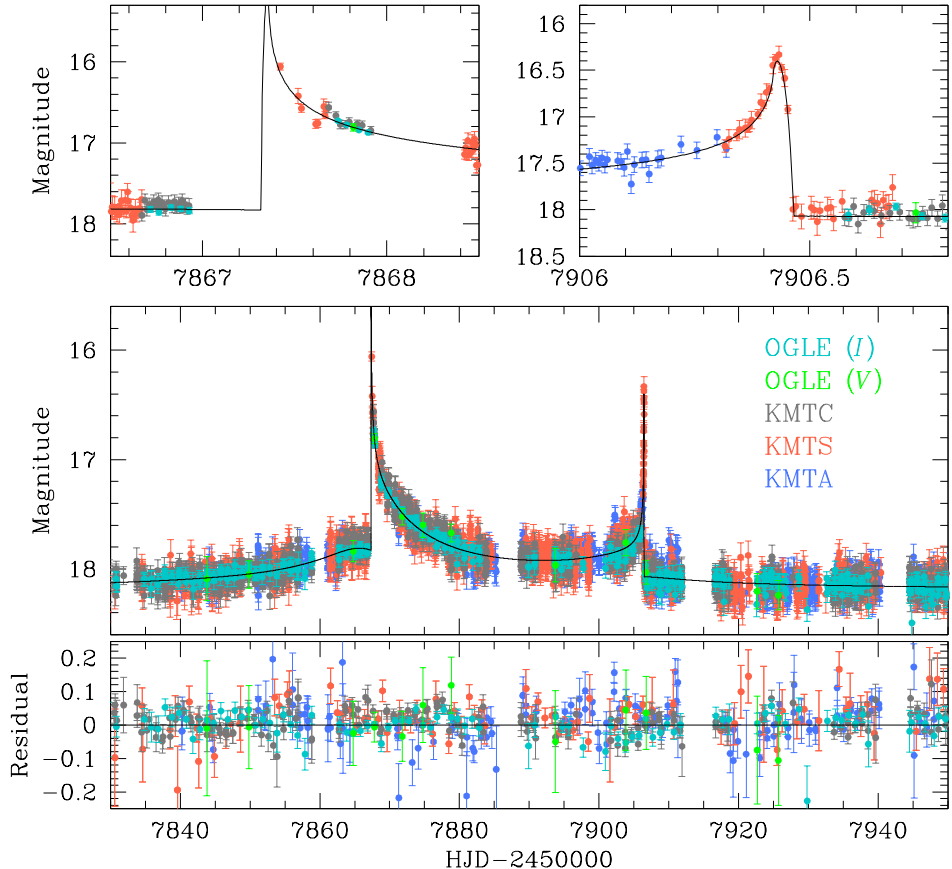}
\caption{
Light curve of OGLE-2017-BLG-0537.
The upper two panels show the zoom of the caustic-crossing parts of the
light curve. The curve plotted over the data points is the model light curve.
The bottom panel shows the residual from the model.
To better show the residual, the data points are binned in one-day bins. 
}
\label{fig:one}
\end{figure*}

Currently, about 3000 microlensing events are annually being detected toward the 
Galactic bulge field by survey experiments (OGLE, MOA, KMTNet). The line of sight 
toward the bulge field passes through the region where the density of stars is 
very high and thus most lensing events are thought to be produced by stellar 
objects \citep{Han2003b}.

Although stars are the major population of lensing objects, it is difficult to 
directly observe them. The most important reason for this difficulty is the slow 
motion of the lens with respect to the lensed star (source).  For typical galactic 
lensing events produced by low-mass stars, the relative lens-source proper motion 
is $\mu\sim 5~{\rm mas}~{\rm yr}^{-1}$. This implies that one has to wait 
$\sim 10$ -- 20 years to resolve the lens from the source even from observations 
using currently available instrument with the highest resolution, e.g., the 
{\it Hubble Space Telescope}  ({\it HST}) with $\sim 0.1^{\prime\prime}$ resolution 
and ground-based Keck adaptive optics (AO) with $\sim 0.05^{\prime\prime}$ 
resolution. In addition, the most common population of lenses 
are low-mass bulge M dwarfs, which have apparent magnitudes fainter than $I\sim 22$. 
Considering that lensing events are detected toward dense fields where stellar 
images are severely blended, it is difficult to observe these faint lens objects.

As an alternate method to reduce the waiting time for the lens-source resolution, 
\citet{Bennett2007} proposed a method using the combination of the ground-based 
data and the data from high-resolution follow-up observations.  In the first step 
of this method, the source flux is estimated from the analysis of the lensing light 
curve obtained from ground-based observations, usually taken in optical bands. In the 
second step, the optical source flux is converted to the photometric system of 
high-resolution data, which is usually taken in near-infrared (NIR) passbands. Then, 
the lens flux is identified as an excess flux measured by subtracting the converted 
source flux from the flux of the target measured in the high-resolution data.  This 
method has been applied to the planetary system OGLE-2007-BLG-368L to estimate the 
mass of the planet host \citep{Sumi2010}.  However, the lens flux estimated by this 
so-called `single-epoch' method is subject to large uncertainties due to the multi-step 
optical-to-NIR flux conversion procedure combined with the difficulty of aligning the 
ground-based and high-resolution images \citep{Henderson2014}.  Furthermore, it is 
difficult to completely rule out companions to the source or lens as well as ambient 
stars as being the origin of excess light, e.g., MOA-2008-BLG-310 \citep{Janczak2010}.

Lens identification from direct imaging has 
been studied and actual observations have been conducted.  Based on the physical and 
dynamical models of the Galaxy, \citet{Han2003a} estimated the fraction of high 
proper-motion events for which the lens and source can be resolved from high-resolution 
observations.  From this, they found that   lenses can be resolved for $\sim 3$ and 22 
per cent of disk-bulge events and for $\sim 0.3$ and 6 per cent of bulge self-lensing 
events from follow-up observations using an instrument with $0.1^{\prime\prime}$ resolving 
power to be conducted 10 and 20 years after events, respectively.  By inspecting events 
detected from 2004 to 2013, \citet{Henderson2014} presented a list of 20 lensing events 
with relative lens-source proper motions greater than $\sim 8~{\rm mas}~{\rm yr}^{-1}$ 
as candidates of follow-up observations for direct lens imaging. For 3 lensing events, 
the lenses were actually detected. These events include MACHO-LMC-5 \citep{Alcock2001}, 
MACHO-95-BLG-37 \citep{Kozlowski2007}, and OGLE-2005-BLG-169 \citep{Batista2015, Bennett2015}.  
Among them, MACHO-LMC-5 was detected toward the Large Magellanic Cloud field and the others 
were detected toward the bulge field. The lenses of the two MACHO events were resolved 
from {\it HST} observations, while the lens of OGLE-2005-BLG-169 was detected from Keck 
AO observations.

In this paper, we present the analysis of the microlensing event OGLE-2017-BLG-0537. 
The event was produced by a binary lens and the light curve of the event exhibits 
characteristic caustic-crossing features.  Despite its very short duration, one of 
the caustic crossings was resolved by high-cadence lensing surveys and we measure 
the relative lens-source proper motion from the analysis of the light curve. It is 
found that the proper motion is $\sim 3$ times bigger than the value of typical galactic 
bulge events. The high proper motion combined with the close distance to the lens 
makes the event a strong candidate for follow-up observations to directly identify the 
lens.

\section{Observations}\label{sec:two}

The source star of the lensing event OGLE-2017-BLG-0537 is located toward the 
Galactic bulge field with the equatorial coordinates 
$({\rm RA},{\rm DEC})_{\rm J2000}=$(17:56:47.75,-28:15:37.8), which correspond to the 
galactic coordinates $(l,b)=(1.83^\circ,-1.76^\circ)$. The baseline magnitude of the 
event before lensing magnification was $I_{\rm base}=18.15$.

The brightening of the source flux induced by gravitational lensing was first noticed 
from observations conducted by the Optical Gravitational Lensing Experiment 
\citep[OGLE:][]{Udalski2015a} survey. The OGLE observations were carried out in $I$ and $V$ 
passbands using the 1.3m telescope located at the Las Campanas Observatory in Chile.

The event was also in the field of the Microlensing Telescope Network 
\citep[KMTNet:][]{Kim2016} survey. The KMTNet observations were conducted using three 
identical 1.6m telescopes that are globally distributed in the Cerro Tololo Interamerican 
Observatory in Chile, the South African Astronomical Observatory in South Africa, and the 
Siding Spring Observatory in Australia. We refer the three KMTNet telescopes as KMTC, KMTS, 
and KMTA, respectively. KMTNet data were taken also in $I$ and $V$ bands. The $V$-band data 
are used mainly to measure the color of the source star.  The event is cataloged 
\citep{Kim2018a,Kim2018b} by KMTNet as BLG02M0506.017562 and lies in two slightly offset 
fields BLG02 and BLG42 with a combined cadence of $\Gamma=4\,{\rm hr}^{-1}$.

Data obtained by the OGLE and KMTNet surveys are processed using the photometry codes 
of the individual groups. Both codes are based on the Difference Image Analysis (DIA) 
developed by \citet{Alard1998} and customized by the individual groups: \citet{Udalski2003} 
and  \citet{Albrow2009} for the OGLE and KMTNet photometry codes, respectively. 
Since the data sets of the individual groups are processed using different photometry 
codes, we renormalize the error bars of the data following the recipe described in 
\citet{Yee2012}.

In Figure~\ref{fig:one}, we present the light curve of the event.  The light curve 
exhibits two strong spikes that are characteristic features of caustic-crossing 
binary-lens events. The caustic crossings occurred at 
${\rm HJD}'={\rm HJD}-2450000\sim 7867.4$ and 7906.4. The facts that the time gap between 
the caustic-crossing spikes, $\sim 39$ days, is very long compared to the duration of the 
event and that the caustic crossings occurred when the lensing magnification was low suggest 
that the projected separation between the lens components is similar to the angular Einstein 
radius. Such a `resonant' binary lens forms a single big closed caustic curve, and the two 
spikes were produced when the source entered and exited the caustic curve. The first caustic 
crossing could have been observed by the KMTA telescope but no observation was conducted due 
to bad weather.  However, the second caustic crossing was captured by the KMTS data thanks 
to the high-cadence coverage of the field. Resolving the caustic crossing enables one to 
measure the normalized source radius $\rho=\theta_*/\thetae$, where $\theta_*$ and $\thetae$ 
represent the angular radii of the source star and the Einstein ring, respectively.  With 
external information about $\theta_*$, one can then measure $\thetae$.

Although the overall light curve has a typical shape of a caustic-crossing binary-lens 
event, we note that the event is very unusual in the sense that the duration of 
caustic-crossing is very short. The caustic-crossing timescale measured during the 
caustic exit is $t_{\rm cc}\sim 0.5$ hr.  The caustic-crossing timescale is related to 
the relative lens-source proper motion by 
\begin{equation}
\mu \simeq {\theta_* \over t_{\rm cc}} {1 \over \sin\psi },
\label{eq1}
\end{equation}
where $\psi$ represents the angle between the source trajectory and the fold of the 
caustic curve. For stars located in the bulge, the angular radius is 
$\theta_*\sim 5~\mu{\rm as}$ for giants and $\sim 0.5~\mu{\rm as}$ for main-sequence 
stars.  Then, the measured caustic-crossing timescale indicates 
$\mu\sim 10~{\rm mas}~{\rm yr}^{-1}$ even for a faint main-sequence source star and 
assuming a right-angle caustic entrance of the source trajectory, i.e., $\psi=90^\circ$. 
This relative lens-source proper motion is significantly greater than those of typical 
lensing events.

\section{Light Curve Modeling}\label{sec:three}

Considering that the two spikes are the characteristic features of binary-lens events, 
we model the observed light curve with a binary-lens interpretation. Under the assumption 
that the relative lens-source motion does not experience any acceleration, i.e., rectilinear 
motion, binary lensing light curves are described by 7 parameters.  Four of these parameters 
describe the lens-source approach, including the time of the closest source-lens approach, 
$t_0$, the impact parameter of the approach, $u_0$, the event timescale, $t_{\rm E}$, and 
the angle between the source trajectory and the binary axis, $\alpha$.  We use the center 
of mass of the binary lens as the reference position on the lens plane, and the impact 
parameter is normalized to the angular Einstein radius.  Another two parameters describe 
the binary lens including the projected separation, $s$ (normalized to $\thetae$), and 
the mass ratio, $q$, between the binary lens components.  Since the event exhibits features 
of caustic crossings, during which the light curve is affected by finite-source effects, 
one needs an additional parameter of the normalized source radius $\rho$ to account for 
these effects.

Due to the large number of lensing parameters, it is difficult to find the best-fit 
solution from a full-scale grid search. We, therefore, utilize a hybrid approach.  In 
this approach, we divide the lensing parameters into two groups of grid $(s, q)$ and 
downhill parameters $(t_0, u_0, t_{\rm E}, \alpha, \rho)$.  This division is based on 
the fact that lensing magnifications are sensitive to the small changes of the grid 
parameters, while magnifications vary smoothly with the change of the downhill parameters.  
With this division of parameters, we conduct a grid search in the space of the grid 
parameters and, for a given set of the grid parameters, we search for the other parameters 
around the circle in the seed values of $\alpha$ using a downhill approach based on the 
Markov Chain Monte Carlo (MCMC) method.  The grid search yields a $\chi^2$ map in the 
space of the grid parameters.  From the map, we identify local minima. In the next step, 
we refine the individual local solutions by allowing all parameters to vary.  We then 
find the global solution by comparing $\chi^2$ values of the individual local minima.

\begin{figure}
\includegraphics[width=\columnwidth]{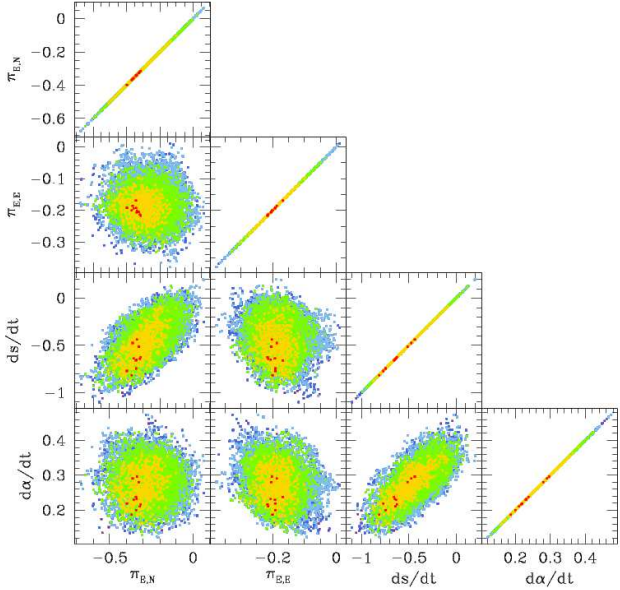}
\caption{
$\Delta\chi^2$ distributions of MCMC chains for the combinations of the higher-order 
lensing parameters $\pien$, $\piee$, $ds/dt$, and $d\alpha/dt$. 
Dots with different color represent chains with $\Delta\chi^2<1$ (red),
$< 4$ (yellow), $<9$ (green), $<16$ (cyan), and $< 25$ (blue).
}
\label{fig:two}
\end{figure}

From the initial search, we find 2 local solutions with event timescales of 
$t_{\rm E}\sim 54$  
days and 90 days, respectively.  Among the two local solutions, we 
exclude the latter solution because (1) it yields a poorer fit than the former solution 
by $\Delta\chi^2\sim 50$ and (2) it requires abnormally large higher-order effects to 
describe the observed light curve while the former solution well describes the observed 
light curve without introducing the higher-order effects.  According to the best-fit 
solution, the event was produced by a binary with a projected separation $s\sim 1.2$ 
and a mass ratio $q\sim 0.4$.

The event lasted $\sim 100$ days, which comprises an important portion of Earth's 
orbital period. In this case, the assumption of the rectilinear lens-source motion 
may not be valid. Two factors can cause non-rectilinear motion. One is the orbital 
motion of Earth, `microlens-parallax' effect \citep{Gould1992}, and the other is the 
orbital motion of the binary lens, `lens-orbital' effect \citep{Albrow2000}. We check 
the possibility of these higher-order effects by conducting additional modeling.  
Accounting for the microlens-parallax effect in lensing modeling requires including 
two additional lensing parameters of $\pi_{{\rm E},N}$ and $\pi_{{\rm E},E}$, which 
denote the two components of the microlens-parallax vector $\pivec_{\rm E}$ along the 
north and east equatorial coordinates, respectively. 
In parallax modeling, there can exist a pair of degenerate solutions with $u_0>0$ 
and $u_0<0$ due to the mirror symmetry of the source trajectory with respect to the 
binary axis: ``ecliptic degeneracy'' \citep{Smith2003, Skowron2011}.  We, therefore, 
consider both $u_0>0$ and $u_0<0$ when parallax effects are considered in modeling.
Considering the lens-orbital 
effect also requires including additional parameters.  Under the approximation that 
the positional changes of the lens components during the event are small, the effect 
is described by two parameters of $ds/dt$ and $d\alpha/dt$, which represent the change 
rates of the binary separation and the source trajectory angle, respectively. We note 
that detecting microlens-parallax effects is important because the physical parameters 
of the lens mass, $M$, and distance to the lens, $D_{\rm L}$, are uniquely determined 
from the measured microlens parallax in combination with the angular Einstein radius by
\begin{equation}
M={\thetae \over \kappa \pie};\qquad
D_{\rm L}={{\rm au} \over \pie\thetae+\pi_{\rm S}},
\label{eq2}
\end{equation}
where $\kappa=4G/(c^2 {\rm au})$, $\pi_{\rm S}={\rm au}/D_{\rm S}$, and $D_{\rm S}$, 
represents the distance to the source.

We find that the higher-order effects in the observed light curve are minor.  It 
is found that the fit improves by $\Delta\chi^2\sim 3.7$ and $\sim 4.1$ with the 
consideration of the microlens-parallax and the lens-orbital effects, respectively.  
When both effects are simultaneously considered, the improvement is $\Delta\chi^2\sim 8.6$.  
In Figure~\ref{fig:two}, we present $\Delta\chi^2$ distributions of MCMC chains 
for the combinations of the higher-order lensing parameters, i.e., 
$\pien$, $\piee$, $ds/dt$, and $d\alpha/dt$. The distributions show no strong 
correlations among the parameters. 
The improvement of the fit, i.e., $\Delta\chi^2\sim 8.6$, mathematically corresponds 
to $\sim 3\sigma$, but such a level of fit improvement can be ascribed to noise in 
data.  We, therefore, cross check the result by comparing the physical parameters 
resulting from the measured $\pie$ with those estimated from Bayesian analysis.  
See detailed discussion in Section~\ref{sec:four}.

\begin{deluxetable}{lcc}
\tablecaption{Best-fit lensing parameters \label{table:one}}
\tablewidth{0pt}
\tabletypesize{\small}
\tablehead{
\multicolumn{1}{c}{Parameter} &
\multicolumn{1}{c}{$u_0>0$}   &
\multicolumn{1}{c}{$u_0<0$}  
}
\startdata                                              
$t_0$   (HJD$^\prime$)     &  7880.795 $\pm$ 0.495     &   7880.684 $\pm$ 0.583     \\   
$u_0$                      &     0.138 $\pm$ 0.007     &     -0.131 $\pm$ 0.008     \\   
$t_{\rm E}$ (days)         &    51.99  $\pm$ 0.61      &     52.21  $\pm$ 0.53      \\   
$s$                        &     1.28  $\pm$ 0.01      &      1.27  $\pm$ 0.01      \\   
$q$                        &     0.50  $\pm$ 0.02      &      0.49  $\pm$ 0.03      \\   
$\alpha$ (rad)             &    -0.595 $\pm$ 0.012     &      0.573 $\pm$ 0.005     \\   
$\rho$  ($10^{-3}$)        &     0.381 $\pm$ 0.020     &      0.389 $\pm$ 0.020     \\   
$\pien$                    &    -0.34  $\pm$ 0.11      &      0.19  $\pm$ 0.10      \\
$\piee$                    &    -0.19  $\pm$ 0.05      &     -0.26  $\pm$ 0.03      \\
$ds/dt$ (yr $^{-1}$)       &    -0.74  $\pm$ 0.19      &     -0.33  $\pm$ 0.18      \\
$d\alpha/dt$ (yr $^{-1}$)  &     0.20  $\pm$ 0.05      &     -0.26  $\pm$ 0.06      \\
$F_s$                      &     0.096 $\pm$ 0.001     &      0.096 $\pm$ 0.001     \\   
$F_b$                      &     0.753 $\pm$ 0.001     &      0.751 $\pm$ 0.001      
\enddata                            
\tablecomments{ 
${\rm HJD}^\prime={\rm HJD}-2450000$.
}
\end{deluxetable}

In Table~\ref{table:one}, we list the lensing parameters of the event. 
Although the higher-order parameters are not securely measured, we present the 
solutions resulting from the models considering the higher-order effects.  It is 
found that the solution with $u_0>0$ is slightly favored over the solution with 
$u_0<0$ by $\Delta\chi^2\sim 3.0$.  We note that the lensing parameters of the 
two solutions with $u_0>0$ and $u_0<0$ are roughly in the relation of 
$(u_0, \alpha, \pien, d\alpha/dt)\leftrightarrow -(u_0, \alpha, \pien, d\alpha/dt)$ 
due to the mirror symmetry of the source trajectories between the two solutions. 
The uncertainties of the individual parameters are estimated from the scatter 
of points in the MCMC chain. Also presented are the fluxes of the source, $F_s$, 
and blend, $F_b$.  
The flux values are normalized to the flux of a star with $I=18$, 
i.e., $F=10^{-0.4(I-18)}$, where $I$ represents the $I$-band magnitude.
The $F_s$ and $F_b$ values show that the blend is $\sim 8$ times brighter than 
the source, indicating that the event is heavily blended.

To be noted among the determined lensing parameter is that the normalized source 
radius $\rho\sim 0.38\times 10^{-3}$ is unusually small. The value of the normalized 
source radius for typical lensing events is $\rho\sim (1-2)\times 10^{-2}$ for a 
giant source star and $\rho\sim (1-2)\times 10^{-3}$ for a main-sequence source 
star.  We find that the source is a main-sequence star located in the bulge (see 
Section~\ref{sec:four_one}).  Then, the measured value of $\rho$ is smaller than 
the value of typical lensing events by a factor $\gtrsim 3$.

\begin{figure}
\includegraphics[width=\columnwidth]{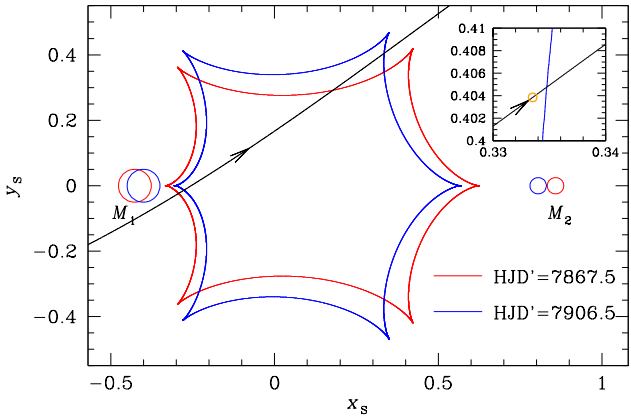}
\caption{
Lens system configuration in which the source trajectory (line
with an arrow) is shown with respect to the caustic (cuspy closed
curve) and the binary lens components (marked by $M_1$ and $M_2$).
We present the lens positions and caustics at two epochs of
${\rm HJD}^\prime=7867.5$ and 7906.5, which correspond to the times of 
the first and second caustic spikes, respectively.
Lengths are scaled to the angular Einstein radius corresponding to the
total mass of the binary lens, and the coordinates are centered at the
center of mass of the lens. The inset shows the zoom of the caustic
region at the time of the source star's caustic exit. The small circle,
whose size is scaled to the source size, indicates the source.
}
\label{fig:three}
\end{figure}

In Figure~\ref{fig:three}, we present the configuration of the lens system in which 
the source trajectory (line with an arrow) with respect to the caustic (cuspy closed 
curve) and the positions of the lens components (marked by $M_1$ and $M_2$) are presented.
We note that the lens positions and the resulting caustic vary in time due to the 
lens-orbital effect.  We, therefore, present the lens positions and caustics at two 
epochs of ${\rm HJD}^\prime=7867.5$ and 7906.5, which correspond to the times of the 
first and second caustic spikes, respectively.
As predicted, the binary lens forms a resonant caustic due to the closeness of the binary 
separation to unity, $s\sim 1.28$.  The caustic forms a closed curve with 6 folds.  
The two spikes in the observed light curve were produced by the source star's crossings 
over the lower left and upper right folds of the caustic. The model light curve of the 
solution and the residual from the model are presented in Figure~\ref{fig:one}.

\section{Constraining the Lens}\label{sec:four}

\subsection{Angular Einstein Radius}\label{sec:four_one}

Measurement of the normalized source radius $\rho$ leads to determination of the 
angular Einstein radius by the relation
\begin{equation}
\thetae={ \theta_*\over \rho}.
\label{eq3}
\end{equation}
For the $\thetae$ determination, then, it is required to estimate the angular source 
radius $\theta_*$.

\begin{figure}
\includegraphics[width=\columnwidth]{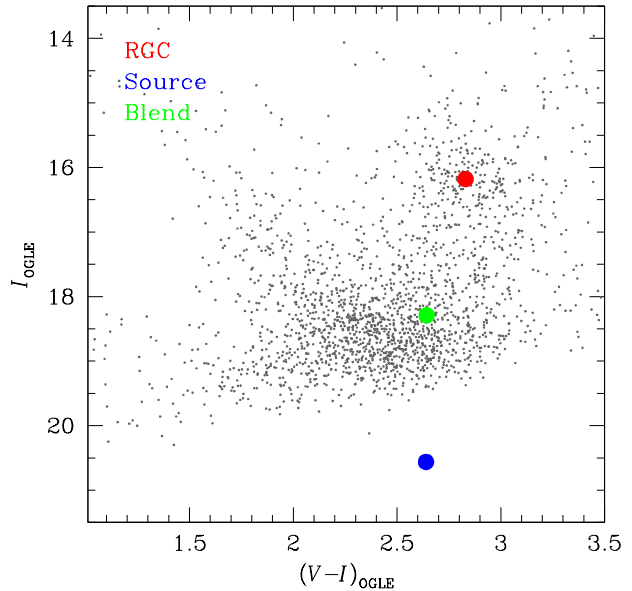}
\caption{
Source location with respect to the centroid of red giant clump (RGC)
in the instrumental color-magnitude diagram. 
Also marked is the position of the blend. 
}
\label{fig:four}
\end{figure}

We estimate the angular radius of the source star based on the color and brightness. 
To measure the de-reddened color $(V-I)_0$ and brightness $I_0$ from the uncalibrated 
color-magnitude diagram (CMD), we use the method described in \citet{Yoo2004}. In this 
method, the centroid of red giant clump (RGC), whose de-reddened color and brightness 
$(V-I,I)_{\rm RGC,0}=(1.06,14.35)$ \citep{Bensby2011, Nataf2013} are known, is used as 
a reference for the calibration of the color and brightness.  Figure~\ref{fig:four} 
shows the source location in the instrumental CMD. 
The instrumental color and magnitude of the source are $(V-I,I)=(2.64 \pm 0.03, 20.56 \pm 0.01)$.  
According to the minimum caustic magnification theorem, the minimum lensing magnification 
when the source is inside caustic is $A_{\rm min}=3$ \citep{Witt1995}.  With this 
constraint combined with the measured brightness at the bottom of U-shape trough 
($I\sim 17.9$ at ${\rm HJD}'\sim 7855$), the upper limit of the source brightness is 
set to be $I_{\rm upper}\sim 19.1$ assuming that all the light comes from the source, 
i.e., no blending.  The measured source brightness, $I\sim 20.6$, is fainter than 
$I_{\rm upper}$ and thus satisfies this constraint.  We measure the source color by 
simultaneously fitting the OGLE $I$ and $V$-band data sets to the best-fit model.  
In Figure~\ref{fig:one}, we plot the OGLE $V$-band data (green dots) used for the 
source color determination.
With the measured offsets in color $(V-I)=-0.19$ 
and brightness $\Delta I=4.38$ of the source with respect to those of the RGC centroid, 
we determine that the de-reddened color and brightness of the source star are 
$(V-I,I)_{{\rm S},0}=(0.87\pm 0.03, 18.73\pm 0.01)$.  This indicates that the source is 
an early K-type main-sequence star. We estimate the angular source radius by converting 
the $V-I$ color into $V-K$ color using the color-color relation of \citet{Bessell1988} 
and then using the $(V-K)/\theta_*$ relation of \citet{Kervella2004}. From this procedure, 
it is estimated that the source has an angular radius of 
$\theta_*=0.68\pm 0.05~\mu{\rm as}$. 
Combined with the measured value of $\rho$, we estimate that the angular Einstein 
radius corresponding to the total mass of the lens is
\begin{equation}
\thetae=1.77\pm 0.16~{\rm mas}.
\label{eq4}
\end{equation}

Once the angular Einstein radius is determined, the relative lens-source proper motion 
is determined with the event timescale by
\begin{equation}
\mu_{\rm geo}={\thetae \over t_{\rm E}}
=12.4 \pm 1.1~{\rm mas}~{\rm yr}^{-1},
\label{eq5}
\end{equation}
and
\begin{equation}
\mu_{\rm hel}=
\left\vert \muvec_{\rm geo} + {\bf v}_{\oplus,\perp}{\pi_{\rm rel}\over {\rm au}}   \right\vert 
= 11.0 \pm 1.0~{\rm mas}~{\rm yr}^{-1},
\label{eq6}
\end{equation}
where $\muvec_{\rm geo}=\mu_{\rm geo}(\pivec_{\rm E}/\pie)$ and $\muvec_{\rm hel}$ represent 
the relative lens-source motion vectors measured in the geocentric and heliocentric frames, 
respectively, and ${\bf v}_{\oplus,\perp}=(-1.6,-21.5)~{\rm km}~{\rm s}^{-1}$ is the velocity 
of the Earth motion projected on the sky at $t_0$.  The position angle of $\muvec_{\rm hel}$ 
is $\sim 196^\circ$ and $\sim 314^\circ$ as measured from the north for the $u_0>0$ and $u_0<0$ 
solutions, respectively.  
In Table~\ref{table:two}, we summarize the values of $\thetae$, $\mu_{\rm geo}$, and 
$\mu_{\rm hel}$ for the $u_0>0$ and $u_0<0$ solutions.
We note that the estimated geocentric proper motion is bigger than the heuristically estimated 
value $\mu\sim 10~{\rm mas}~{\rm yr}^{-1}$ in Section~\ref{sec:two} because the caustic entrance 
angle of the source trajectory $\psi \sim 53^\circ$ is different from 
the assumption of the right-angle entrance.  See the zoom of caustic region at the moment 
of the source star's caustic exit presented in the inset of Figure~\ref{fig:three}.

We note that the estimated angular Einstein radius is significantly bigger than the 
values of typical lensing events. For the most common population of galactic bulge 
events (with $D_{\rm L}\sim 6$ kpc) produced by low-mass stars (with $M\sim 0.4~M_\odot$), 
the angular Einstein radius is $\thetae\sim 0.4$ mas.  Then, the measured angular 
Einstein radius is $\sim 4.4$ times bigger than the typical value. The angular Einstein 
radius is related to the physical lens parameters by
\begin{equation}
\thetae=\sqrt{\kappa M \pi_{\rm rel}};\qquad
\pi_{\rm rel}={\rm au}\left( {1\over D_{\rm L}} - {1\over D_{\rm S}}\right).
\label{eq7}
\end{equation}
Since $\thetae\propto M^{1/2}$ and $\thetae\propto \pi_{\rm rel}^{-1/2}$, the large 
value of $\thetae$ indicates that the lens is either very heavy or located close to 
the observer. The large angular Einstein radius leads to the high relative lens-source 
proper motion because $\mu\propto \thetae$.  As we will discuss in the following subsection, 
the high proper motion makes the event an important target for high-resolution follow-up 
observations for direct detection of the lens.

\begin{figure*}
\epsscale{0.85}
\plotone{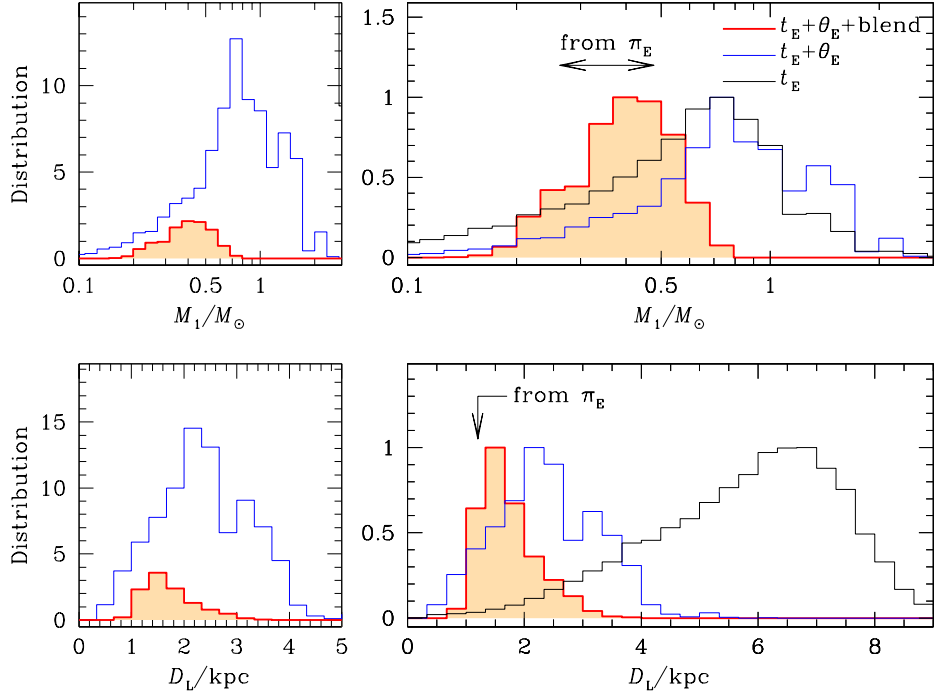}
\caption{
The probability distributions of the primary mass (upper panels) and distance to the 
lens (lower panels) obtained by Bayesian analysis.  The 3 curves in black, blue, and 
red colors represent the distributions obtained with only the event timescale constraint, 
with timescale plus angular Einstein radius constraint, and with the additional constraint 
of the blended light, respectively.  The distributions in the left panels are relatively 
scaled, while the distributions in the right panels are scaled so that the peak of the 
distribution becomes unity.
The lens mass and distance marked by ``from $\pie$'' represent the values estimated 
from the measured microlens parallax. 
}
\label{fig:five}
\end{figure*}


\subsection{Physical Lens Parameters: Based on $\pie$}\label{sec:four_two}

Although the measured microlens parallax $\pie$ is subject to some uncertainty due to 
the weak parallax signal, we estimate the physical lens parameters based on the parallax 
parameters combined with the measured angular Einstein radius using the relations in 
Equation~(\ref{eq2}).  In Table~\ref{table:two}, we present the estimated parameters.  
Here $M_1$ and $M_2$ ($< M_1$) represent the masses of the individual lens components.  
The projected separation between the lens components is computed by 
$a_\perp=sD_{\rm L}\thetae$.  According to the estimated masses of the lens components, 
the lens consists of an early and a mid M-type dwarfs with masses 
$M_1\sim 0.4~M_\odot$ and $M_2\sim 0.2~M_\odot$, respectively. 
As anticipated by the large 
angular Einstein radius, it is estimated that the lens is located at a close distance 
of $D_{\rm L}\sim 1.2$ -- 1.5 kpc.

Also presented in Table~\ref{table:two} is the projected kinetic-to-potential energy 
ratio, (KE/PE)$_\perp$.  The ratio is computed based on the measured lensing parameters 
of $s$, $ds/dt$, and $d\alpha/dt$ by
\begin{equation}
\left( {{\rm KE} \over {\rm PE}}\right)_\perp = 
{ (a_\perp/{\rm au})^3 \over 8\pi(M/M_\odot)}
\left[ \left( {1\over s}{ds/dt\over {\rm yr}^{-1}}\right)^2 + 
\left( {d\alpha/dt\over {\rm yr}^{-1}} \right)^2\right].
\label{eq8}
\end{equation}
In order for the lens to be a bound system, the ratio should be less than unity.  It 
is found that both $u_0>0$ and $u_0<0$ solutions meet this requirement.

\begin{deluxetable}{lccc}
\tablecaption{Physical Lens Parameters (from $\pie$) \label{table:two}}
\tablewidth{0pt}
\tabletypesize{\small}
\tablehead{
\multicolumn{1}{c}{Parameter} &
\multicolumn{1}{c}{$u_0>0$}   &
\multicolumn{1}{c}{$u_0<0$}   &
\multicolumn{1}{c}{Bayesian}  
}
\startdata                                              
$\thetae$    (mas)              & $1.77 \pm 0.16$    &   $1.73 \pm 0.16$  & -                     \\
$\mu_{\rm geo}$ (mas~yr$^{-1}$) & $12.4 \pm 1.1$     &   $12.1 \pm 1.1$   & -                     \\
$\mu_{\rm hel}$ (mas~yr$^{-1}$) & $11.0 \pm 1.0$     &   $10.3 \pm 0.9$   & -                     \\
$M_1$        ($M_\odot$)        & $0.37 \pm 0.10$    &   $0.44 \pm 0.11$  &  $0.43 \pm 0.14$      \\
$M_2$        ($M_\odot$)        & $0.18 \pm 0.05$    &   $0.22 \pm 0.06$  &  $0.22 \pm 0.07$      \\
$D_{\rm L}$  (kpc)              & $1.2  \pm 0.3$     &   $1.5  \pm 0.3$   &  $1.8^{+0.6}_{-0.4}$  \\
$a_\perp$    (au)               & $2.8  \pm 0.6$     &   $3.2  \pm 0.7$   &  $3.5^{+1.2}_{-0.8}$  \\
(KE/PE)$_\perp$                 & 0.18               &   0.09             & -             
\enddata                            
\end{deluxetable}

\subsection{Physical Lens Parameters: Bayesian Approach}\label{sec:four_three}

The lens parameters determined in the previous subsection are not very secure due 
to the uncertain microlens parallax value.  We, therefore, cross check the physical 
parameters by additionally conducting a Bayesian analysis of the event based on the measured 
event timescale and the angular Einstein radius combined with the blended light.  
We note that the blended light provides a constraint because the lens cannot be 
brighter than the blend.

For the Bayesian analysis, one needs models of the mass function and the physical 
and dynamical distributions of lens and source.  In the analysis, we use the 
\citet{Chabrier2003} mass function. 
In the mass function, we include stellar remnants.  Following the model of 
\citet{Gould2000}, we assume that stars with masses 
$1~M_\odot\leq M \leq 8~M_\odot$, 
$8~M_\odot\leq M \leq 40~M_\odot$,
$M \geq 40~M_\odot$,
have evolved into white dwarfs (with a mean mass $\langle M\rangle \sim 0.6~M_\odot$), 
neutron stars ($\langle M\rangle \sim 1.36~M_\odot$), black holes 
($\langle M\rangle \sim 5~M_\odot$), respectively.  
For the physical distributions of lens objects, we use the density model of 
\citet{Han2003b}, in which the Galaxy is described by a double-exponential disk 
and a triaxial bulge. To describe the motion of the lens and source, we use the 
dynamical model of \citet{Han1995}, in which disk objects move following a gaussian 
distribution with a mean corresponding to the disk rotation speed and bulge objects 
move according to a triaxial Gaussian distribution with velocity components 
determined by tensor virial theorem based on the bulge shape.

With the adopted model distributions, we produce a large number of mock lensing 
events by conducting Monte Carlo simulation in which the mass of the primary lens 
is drawn from the model mass function.  We then construct the distributions 
of the lens mass and distance for events with lens masses and distances fall in the 
ranges of the measured event time scale and angular Einstein radius, and with lens 
brightness fainter than the blend.  
We note that the constraints of the time scale and angular Einstein radius are 
based on the values corresponding to the primary of the lens, i.e., 
$t_{{\rm E},1}=t_{\rm E}/\sqrt{1+q}$ and $\theta_{{\rm E},1}=\theta_{\rm E}/\sqrt{1+q}$, 
under the assumption that binary components follow the same mass function as that 
of single stars.  
Then, the mass and distance are estimated as the median values and their 
uncertainties are estimated based on the 16/84 percentiles of the distributions.

Figure~\ref{fig:five} shows the probability distributions of the primary mass 
(upper panels) and distance to the lens (lower panels).  To show the how the 
individual constraints, i.e., $\te$, $\thetae$, and blended light, contribute to 
characterize the physical lens parameters,  we present 3 distributions: with only 
the event timescale constraint (marked by ``$t_{\rm E}$''), with the timescale 
plus angular Einstein radius constraint (``$t_{\rm E}+\theta_{\rm E}$''), and with 
the additional constraint of the blended light (``$t_{\rm E}+\theta_{\rm E}+$ blend'').  
We note that the distributions in the left panels are relatively scaled, while the 
distributions in the right panels are scaled so that the peak of the distribution 
becomes unity.  From the comparison of the distributions, it is found that the 
angular Einstein radius gives a strong constraint on the distance to the lens.  
This is because 
$\thetae \propto \pi_{\rm rel}^{1/2} \propto (D_{\rm L}^{-1}-D_{\rm S}^{-1})^{1/2}$ 
and thus nearby lenses can result in large angular Einstein radius.  On the other 
hand, the blended light provides an important constraint on the lens mass.  
Although heavy lenses (with $M_1\gtrsim 0.8\ M_\odot$) can produce events with 
large $\thetae$, they are excluded by the constraint of the lens brightness.  
The blend constraint also excludes very nearby lenses (with $\dl \lesssim 1$ kpc).

The masses of the primary and the companion estimated by the Bayesian analysis are
\begin{equation}
M_1 = 0.43 \pm 0.14\ M_\odot
\label{eq9}
\end{equation}
and
\begin{equation}
M_2 = 0.22\pm 0.07\ M_\odot.
\label{eq10}
\end{equation}
We note that the lens masses span wide ranges, 
$0.29 \lesssim M_1/M_\odot \lesssim 0.57$ and $0.15 \lesssim M_2/M_\odot \lesssim 0.29$
(as measured in $1\sigma$ level), due to the nature 
of the Bayesian mass estimation.  The lens components are separated in projection by
\begin{equation}
a_\perp=3.5_{-0.8}^{+1.2}\ {\rm au}.
\label{eq11}
\end{equation}
The estimated distance to the lens is
\begin{equation}
D_{\rm L}=1.8_{-0.4}^{+0.6}\ {\rm kpc},
\label{eq12}
\end{equation}
indicating that the lens is located at a close distance.
We list the physical lens parameters estimated from the Bayesian analysis in Table~\ref{table:two}.

We note that the physical lens parameters estimated based on the measured microlens parallax 
match well those estimated from the Bayesian analysis.
In Figure~\ref{fig:five}, we mark the lens mass and distance estimated from $\pie$
on the probability distributions obtained from the Bayesian analysis.
It is found that both the lens mass and distance are positioned 
close to the highest probability regions of the Bayesian distributions.
This indicates that the microlens parallax is correctly measured despite the weak 
signal in the lensing light curve.

\section{Direct Lens Imaging}\label{sec:five}

The scientific importance of the event lies in the fact that the event provides a 
rare opportunity to test the microlensing model from follow-up observations.  The 
main reason for the event to be a good target for checking the lensing parameters 
is that the lens can be resolved from the source and thus directly imaged within a 
reasonably short period of time after the event.  The measured proper motion, 
$\mu =12.4\pm 1.0~{\rm mas}~{\rm yr}^{-1}$, of the event is the third highest value 
among the events with measured proper motions after OGLE-2007-BLG-224, with 
$\mu =48\pm 2~{\rm mas}~{\rm yr}^{-1}$ \citep{Gould2009}, and LMC-5, with 
$\mu =21.4\pm 0.7~{\rm mas}~{\rm yr}^{-1}$ \citep{Alcock2001, Drake2004, Gould2004}.  
We also note that the event has the second-largest well-measured angular Einstein 
radius of published lensing events 
after OGLE-2011-BLG-0417 \citep{Shin2012}.  See \citet{Henderson2014} and 
\citet{Penny2016} for the compilation of events with well measured angular Einstein radii.
\citet{Batista2015} pointed out that a lens 
can be resolved from a source on Keck images when the separation between the lens 
and source is $\sim 50$ -- 60~mas.  Applying the same criterion, then, the lens of 
OGLE-2017-BLG-0537 can be resolved from Keck AO imaging if follow-up observations 
are conducted $\gtrsim 4$ years after the event.

\begin{figure}
\includegraphics[width=\columnwidth]{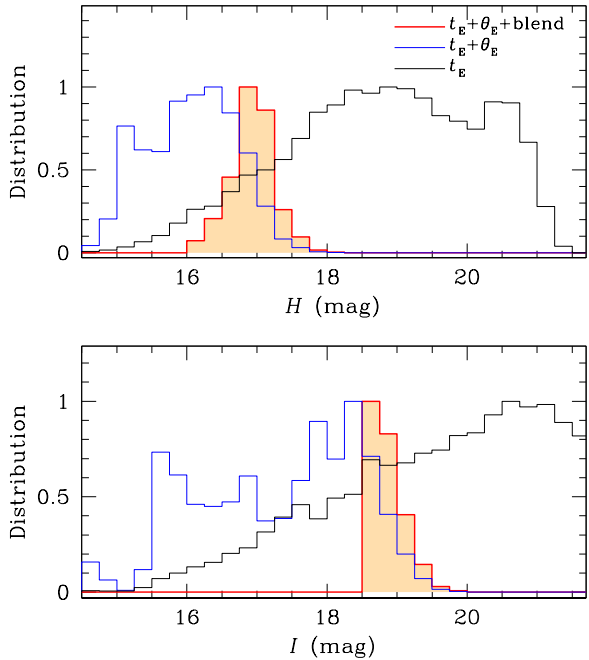}
\caption{
The probability distributions of the $H$ and $I$-band brightness of the lens. 
Notations are same as those in Figure~\ref{fig:five}.
}
\label{fig:six}
\end{figure}

Another reason that makes the event favorable for high-resolution follow-up observations 
is the closeness of the lens.  The most common population of lenses are low-mass stars. 
Then, if an event is produced by such a lens located at a large distance, e.g., bulge 
lens, it would be difficult to image the lens.  In the case of OGLE-2017-BLG-0537, the 
lens will be bright enough for direct imaging because of its small distance despite that 
the lens components are M dwarfs.  
In Figure~\ref{fig:six}, we present the distribution of the lens brightness estimated 
from the Bayesian analysis. 
We estimate $H$-band brightness because follow-up observations will be 
conducted in NIR bands. In the Bayesian simulation, we estimate $H$-band magnitude 
using the relation
\begin{equation}
H=M_V - (V-H) + 5\log D_{\rm L}/{\rm pc} - 5 + A_H.
\label{eq13}
\end{equation}
Here we estimate the absolute $V$-band magnitude, $M_V$, from the mass using the 
mass-luminosity relation and use $V-H$ color 
from from \citet{Bessell1988}. We assume $A_H \sim 0.108 A_V$ \citep{Nishiyama2008},
where $A_V\sim 3.3$ is obtained from the OGLE extinction map. 
The estimated range of the $H$-band brightness of the lens is
\begin{equation}
H = 17.1_{-0.33}^{+0.28}.
\label{eq14}
\end{equation}
In comparison, the $H$-band source brightness is $H_{\rm S}\sim 18.0$.

We note that the apparent brightness of the blend, $I_b\sim 18.3$, matches well the 
lens brightness estimated by Bayesian analysis, $I_{\rm L}\sim 18.5$. This indicates 
that the blend light is likely to come from the lens.  Under this assumption, 
prompt high-resolution imaging can potentially constrain the lens properties 
without needing to wait.  See \citet{Henderson2015} for detailed discussion.

There exist several cases in which lensing models are checked by follow-up observations. 
The first example is OGLE-2005-BLG-169 in which a weak and short-term anomaly in the 
lensing light curve was produced by a lens with a Neptune mass ratio planetary companion 
\citep{Gould2006}.  From the Keck AO observation conducted $\sim 8$ years after the 
discovery of the event, the lens and source were completely resolved, proving a precise 
measurement of the relative lens-source motion \citep{Batista2015}.  This confirmed and 
refined the lensing model and ruled out a range of solutions that were allowed by the 
microlensing light curve.  The lensing solution of the event was additionally confirmed 
from follow-up observations conducted using {\it HST} \citep{Bennett2015}.  Another case 
where lensing solution was checked by follow-up observations is the binary event 
OGLE-2011-BLG-0417.  Microlensing analysis of the event yielded not only the lens mass 
and distance but also a complete Keplerian solutions including the radial velocity of 
the binary lens orbital motion \citep{Shin2012}.  This led to the proposal of follow-up 
spectroscopic observations for Doppler radial velocity measurement to test the lensing 
solution \citep{Gould2013}.  \citet{Boisse2015} actually conducted the proposed 
radial-velocity measurement of the event using UVES spectroscopy mounted on the VLT, 
but this measurement did not confirm the microlensing prediction for the binary lens 
system.  Using the spectral energy distribution combined with the high-resolution images 
obtained from successive follow-up spectroscopic and AO observations of the event, 
\citet{Santerne2016} concluded that the lens parameters were not compatible with the 
ones obtained from lensing analysis, raising the need to check the lensing solution.  
Additionally, the lensing parameters of the planetary event OGLE-2014-BLG-0124 
determined by \citet{Udalski2015b} were confirmed and refined by follow-up AO imaging 
observation conducted by \citet{Beaulieu2018}.

\section{Conclusion}\label{sec:six}

We analyzed the microlensing event OGLE-2017-BLG-0537, in which the light curve exhibited 
two strong caustic-crossing spikes.  It was found that the event was produced by a binary 
lens for which the lens components with a mass ratio $q\sim 0.5$ was separated in projection 
by $s\sim 1.3$.  Among the two caustic-crossing spikes, the second caustic crossing 
was resolved by high-cadence surveys.  Analysis of the caustic-crossing part of the light 
curve yielded an angular Einstein radius of $\thetae=1.77\pm 0.16$ mas and a lens-source 
relative proper motion of $\mu =12.4\pm 1.1~{\rm mas}~{\rm yr}^{-1}$.  The measured proper 
motion was the third highest value among the events with measured proper motions and 
$\sim 3$ times higher than the value of typical galactic bulge events.  This makes the event 
a strong candidate for follow-up observations to directly image the lens from the source.

From the angular Einstein radius combined with the microlens parallax, it was estimated 
that the lens was composed of two main-sequence stars with masses $M_1\sim 0.4~M_\odot$ 
and $M_2\sim 0.2~M_\odot$ located at a distance of $D_{\rm L}\sim 1.2$~kpc, but  
the physical lens parameters were not very secure due to the weak microlens-parallax 
signal.
We cross checked the physical parameters by additionally conducting a Bayesian analysis 
based on the measured Einstein radius and event timescale combined with the blending 
constraint.  From this, we found that the physical parameters estimated from the Bayesian 
analysis were consistent with those based on the measured microlens parallax.  
Resolving the lens from the source can be done in about 5 years from high-resolution follow-up 
observations and this will provide a rare opportunity to test the microlensing model
and refine lens parameters.

\acknowledgments
Work by C.~Han was supported by the grant (2017R1A4A1015178) of
National Research Foundation of Korea.
The OGLE project has received funding from the National Science Centre, Poland, grant 
MAESTRO 2014/14/A/ST9/00121 to A.~Udalski.
Work by WZ, YKJ, and AG were supported by AST-1516842 from the US NSF.
WZ, IGS, and AG were supported by JPL grant 1500811.
Work by YS was supported by an appointment to the NASA Postdoctoral Program at the Jet
Propulsion Laboratory, administered by Universities Space Research Association
through a contract with NASA.
This research has made use of the KMTNet system operated by the Korea
Astronomy and Space Science Institute (KASI) and the data were obtained at
three host sites of CTIO in Chile, SAAO in South Africa, and SSO in
Australia.
We acknowledge the high-speed internet service (KREONET)
provided by Korea Institute of Science and Technology Information (KISTI).

\end{document}